\documentstyle[epsfig,12pt]{article}
\begin{document} 
%
%
%
\newcommand{\x}{\cdot}
\newcommand{\ra}{\rightarrow}
\newcommand{\pom}{\mbox{$\rm{\cal P}$omeron}}
\newcommand{\flux}{\mbox{$ F_{{\cal P}/p}(t, \xi)$}}
\newcommand{\ap}{\mbox{$\bar{p}$}}
\newcommand{\pap}{\mbox{$p \bar{p}$}}
\newcommand{\SPS}{\mbox{S\pap S}}
\newcommand{\xp}{\mbox{$x_{p}$}}
\newcommand{\et}{\mbox{${E_T}$}}
\newcommand{\etj}{\mbox{$\et ^{jet}$}}
\newcommand{\sumet}{\mbox{$\Sigma \et$}}
\newcommand{\sumetj}{\mbox{$\Sigma \et ^{jet}$}}
\newcommand{\pt}{\mbox{${p_T}$}}
\newcommand{\mpr}{\mbox{${ m_p}$}}
\newcommand{\mpi}{\mbox{${ m_\pi}$}}
\newcommand{\rs}{\mbox{$ \sqrt{s}$}}
\newcommand{\rsp}{\mbox{$ \sqrt{s'}$}}
\newcommand{\rsps}{\mbox{$ \sqrt{s} = 630 $ GeV}}
\newcommand{\lum}{\mbox{$\int {\cal L} {\rm dt}$}}
\newcommand{\T}{\mbox{$t$}}
\newcommand{\abt}{\mbox{${ |t|}$}}
\newcommand{\di}{\mbox{d}}
\newcommand{\sigdifjets}{\mbox{$ \sigma_{sd}^{jets}$}}
\newcommand{\sigpomjets}{\mbox{$ \sigma_{{\cal P}p}^{jets}$}}
\newcommand{\sigdiftot}{\mbox{$ \sigma_{sd}^{total}$}}
\newcommand{\sigpomtot}{\mbox{$ \sigma_{{\cal P}p}^{total}$}}
\newcommand{\dsig}   {\mbox{$ {{ d^2 \sigma        }\over{d \xi dt}} $}}
\newcommand{\dsigdif}{\mbox{$ {{ d^2 \sigma _{sd}  }\over{d \xi dt}} $}}

\newcommand{\alamb}{\mbox{$\overline{\Lambda^{\circ}}$}}
\newcommand{\lamb}{\mbox{$\Lambda^{\circ}$}} 
\newcommand{\PRET}{\mbox{\Proton-\sumet}}
\begin{titlepage}
\begin{center}
{\large   }
{\large EUROPEAN ORGANIZATION FOR NUCLEAR RESEARCH}
\end{center}
\vspace{2 ex}
\begin{flushright}  
{
31 December, 1997
}
\end{flushright}
 
  
\begin{center}
{
\LARGE\bf
\rule{0mm}{7mm} Cross Section Measurements of\\
\rule{0mm}{7mm} Hard Diffraction at the {\boldmath \SPS}-Collider
}\\
\vspace{4ex}

A. Brandt$^{1a}$, S. Erhan$^{b}$, A. Kuzucu$^{2}$, M. Medinnis$^{3}$,\\
N. Ozdes$^{2}$, P.E. Schlein, M.T. Zeyrek$^{4}$, J.G. Zweizig$^{5}$\\
University of California$^{*}$, Los Angeles, California 90024, USA. \\
\vspace{2 ex}
J.B. Cheze, J. Zsembery \\
Centre d'Etudes Nucleaires-Saclay, 91191 Gif-sur-Yvette, France.
\end{center}
\vspace{2 ex}

\centerline{(UA8 Collaboration)}

\vspace{1 ex}

\begin{abstract}

The UA8 experiment previously
reported the observation of jets in diffractive events containing leading 
protons (``hard diffraction"), 
which was interpreted as evidence for the partonic structure of 
an exchanged Reggeon, believed to be the \pom .
In the present Letter, we report the final UA8 hard-diffractive (jet)
cross section results and their interpretation.
After corrections, the fraction of single diffractive events 
with mass from 118 to 189~GeV that have two scattered partons, each with 
$\etj > 8$~GeV, is in the range 0.002 to 0.003 (depending on \xp ).
We determine the product, $fK$, of
the fraction by which the \pom 's momentum sum rule is violated and
the normalization constant of the \pom -Flux-Factor of the 
proton. 
For a pure gluonic- or a pure $q\bar{q}$-\pom , respectively:
$fK$ = ($0.30 \pm 0.05 \pm 0.09$) and  ($0.56 \pm 0.09 \pm 0.17$) 
GeV$^{-2}$.
\end{abstract}

\begin{center}
Submitted to Physics Letters B \\
\end{center}
\vspace{1 ex}
\rule[.5ex]{16cm}{.02cm}
$^{*}$ Supported by U.S. National Science Foundation
Grant PHY94-23142 \\
$^{a}$ email:  brandta@fnalv.fnal.gov \\
$^{b}$ email:  samim.erhan@cern.ch \\
$^{1}$ Now at Fermi National Accelerator Laboratory, Batavia, Illinois, 
U.S.A. \\ 
$^{2}$ Visitor from Cukurova University, Adana, Turkey; also supported by 
ICSC - World Lab.\\
$^{3}$ Present address: DESY, Zeuthen, Germany \\
$^{4}$ Visitor from Middle East Tech. Univ., Ankara, Turkey; supported by Tubitak. \\
$^{5}$ Present address: DESY, Hamburg, Germany  \\
\end{titlepage}      

\pagebreak
\setlength{\oddsidemargin}{0 cm}
\setlength{\evensidemargin}{0 cm}
\setlength{\topmargin}{0.5 cm}
\setlength{\textheight}{22 cm}
\setlength{\textwidth}{16 cm}
\setcounter{totalnumber}{20}
\clearpage\mbox{}\clearpage
\pagestyle{plain}
\setcounter{page}{1}


\section{Introduction}
\label{intro}

\indent

During the last decade, the physics of \pom -exchange
or diffractive (leading proton) processes,
\begin{equation}
\bar{p} \, + \, p_i \, \ra \, X \,+\, p_f \, \, \, \, \, \, \, \, \, \, \, \, 
\, \, \, + \, \, \,  charge \, \, conjugate,
\label{eq:ppdif}
\end{equation}
\begin{equation}
e^{\mp} \, + \, p_i \, \ra \, e^{\mp} \, + \, X \, + \, p_f, 
\, \, \, \, \, \, \, \, \, \, \, \, \, \, \, \, 
\label{eq:epdif}
\end{equation}
\noindent
has taken a new direction:

\begin{itemize}

\item Ingelman and Schlein\cite{is} proposed that the partonic structure
of the exchanged Reggeons in Reactions~\ref{eq:ppdif} and \ref{eq:epdif}
(dominated by the \pom  \cite{goul})
could be studied if hard-scattering effects were observed
in the interactions of the exchanged Reggeon with the $\bar{p}$ in the
first process and with a photon in the second. 
Based on the assumption of factorization, a method of analysis was
proposed to extract the \pom\ structure function.
\item This experiment, UA8 at the CERN collider (\rs\ = 630 GeV), 
presented the first evidence\cite{bonino} 
that the \pom\ has a partonic structure, with the observation of
QCD jet production in React.~\ref{eq:ppdif}.
The observed event rate had the predicted\cite{is} order of magnitude
from \pom\ phenomenology.
In a second Letter\cite{brandt}, which reported a sample of 300 2-jet events
with $\etj > 8$~GeV, 
an analysis of the longitudinal momentum distribution of the 2-jet system
in the \pom -proton frame showed that 
the \pom\ internal structure is ``hard'', like $x(1-x)^1$, with 
about 30\% of the sample exhibiting a $\delta$-function-like structure near 
$x = 1$. 
Furthermore, the fraction of diffractive events which exhibit hard scattering 
was observed to be independent of momentum-transfer, $|\T|$, over the range
0.8-2.0 GeV$^2$.
\item The ZEUS\cite{zeus} and H1\cite{h1} 
experiments at HERA have observed related Deep-Inelastic hard-diffraction
events in React.~\ref{eq:epdif}. 
They also find evidence for a hard \pom\ structure but, in addition,
are able to demonstrate that there is a large gluonic component.
In particular, H1 has recently presented\cite{h1gluonic} a 
QCD analysis of their data, from which they 
conclude that gluons carry 80--90\% of
the \pom 's momentum and that, at small--Q$^2$, there is a parton
concentration near $x=1$ in the \pom\ system. 
This observation may be intimately related to the
``super-hard" \pom\ structure reported by us in Ref.~\cite{brandt}.

The D\O\ collaboration has confirmed the existence of hard diffraction
in \pap\ interactions at \rs\ = 630 GeV and also report its existence at 
\rs\ = 1800 GeV\cite{d0pp96}.
At 1800 GeV, the CDF collaboration has also 
obtained evidence that the \pom\ is dominantly
gluonic, by comparing the measured rates of diffractive 
W-boson\cite{cdfw} and dijet\cite{cdfdijet} production.

\end{itemize}

Since the UA8 jet analyses 
probed the structure of the $\xi = 1 - \xp $ component of the proton,
{\it independent of any assumptions about its identity},
it is important to study the jet cross section within the context of 
\pom\ phenomology.
In this Letter, we report the final UA8 hard diffractive (jet) cross section
results and their interpretation. We extract new parameters from the data which 
can be used to predict other hard-diffraction cross sections, thus allowing 
tests of factorization and other aspects of hard diffraction phenomenology. 

Preliminary results from these analyses were presented\cite{marseille} at 
the 1993 Marseille Conference.
Since then, much work has been done to further understand the phenomenology of
single diffraction and the \pom -Flux-Factor, which is necessary for a more
thorough understanding of the data.
In particular, a detailed analysis\cite{ua8dif} of our UA8 
data, together with the data from other experiments has been performed. Some of
the relevant results are discussed below.

We attempt to clarify several items.
One key issue is to what extent the \pom\ behaves 
like a real particle, in the
sense that the momentum fractions of its partons sum to unity
(the ``momentum sum rule")\cite{dl_hard,berger}.
Another has to do with the (arbitrary) conventions used for 
the normalization of the \pom -Flux-Factor in the proton, 
an overall scale for the process,
for which at least three versions exist in the 
literature\cite{dl_hard,berger,dino}.


\section{Diffractive Jet Data sample}
\label{cross}

\indent

The momentum of the final state proton in Reaction~\ref{eq:ppdif},
$p_f$, was measured in one of four 
small-angle UA8 ``Roman-pot" spectrometers\cite{ua8hard}
which were interfaced to the UA2 experiment\cite{ua2};
the final-state jets were measured in the upgraded UA2 calorimeter 
system.
The inclusive proton data sample was provided by the so-called ``DIF''
trigger, whose data-acquisition logic required a proton or antiproton 
with an acceptable momentum that was calculated online\cite{ua8hard,jgz}. 
A second trigger, used to provide the jet event sample and denoted ``JET'',
combined the DIF trigger with the additional requirement that the total 
transverse energy in the UA2 calorimeter system had $ \sumet > 18$ GeV
(this cut was increased to 22 GeV in the offline analysis).

In Reaction~\ref{eq:ppdif}, the incident \ap\ interacts with a residual 
component of the incident proton, $p_i$, with beam 
momentum fraction\footnote{Because $\xp + \xi = 1$, we may refer to one
or the other of these equivalent variables in this Letter.}, 
$\xi = 1 - \xp $, where $\xp = p_f / p_i$. 
The system $X$ in Reaction~\ref{eq:ppdif} 
has squared invariant mass, $s' = s \xi $, so that in this experiment, 
for example, \rsp\ = 118 (200) GeV when $\xi$ = 0.035 (0.10). 

\begin{figure}
\begin{center}
\mbox{\epsfig{file=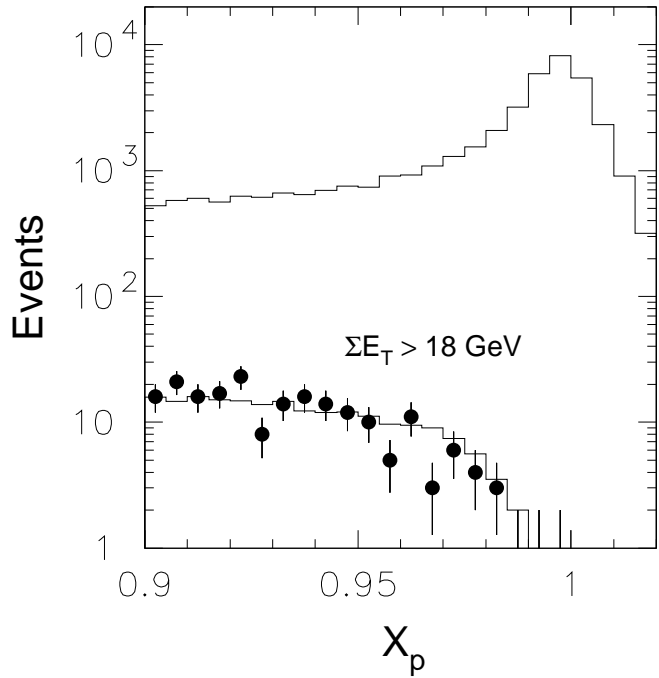,width=16cm}}
\end{center}
\caption[]{
The upper histogram shows the observed dependence of event yield on proton 
momentum fraction, \xp , for ``DIF'' trigger events (inclusive
protons or antiprotons). 
The solid points are those ``DIF'' trigger events
that have $\sumet > 18$ GeV (offline evaluation); 
The lower histogram, which is
normalized to the solid points, corresponds to the high-statistics sample for 
which the same \sumet\ selection was imposed online in the ``JET'' trigger.
}
\label{fig:xp}
\end{figure}

Figure~\ref{fig:xp} shows our observed inclusive proton \xp\ distribution
for both triggers. 
For the DIF trigger, the most likely value of \xp\ is near
unity and, correspondingly, the most likely value of $\xi$ is near 
zero. 
On the same plot, the solid points are those DIF-trigger events which
satisfy the offline requirement, $ \sumet > 18$ GeV. The lower
histogram which is normalized to the solid points corresponds to the
high-statistics sample for which the same \sumet\ selection
was imposed online in the JET trigger. 
The \sumet\ selection discriminates against \xp\ values near 1.0, which
are incapable of producing large $s'$ values.

\begin{figure}
\begin{center}
\mbox{\epsfig{file=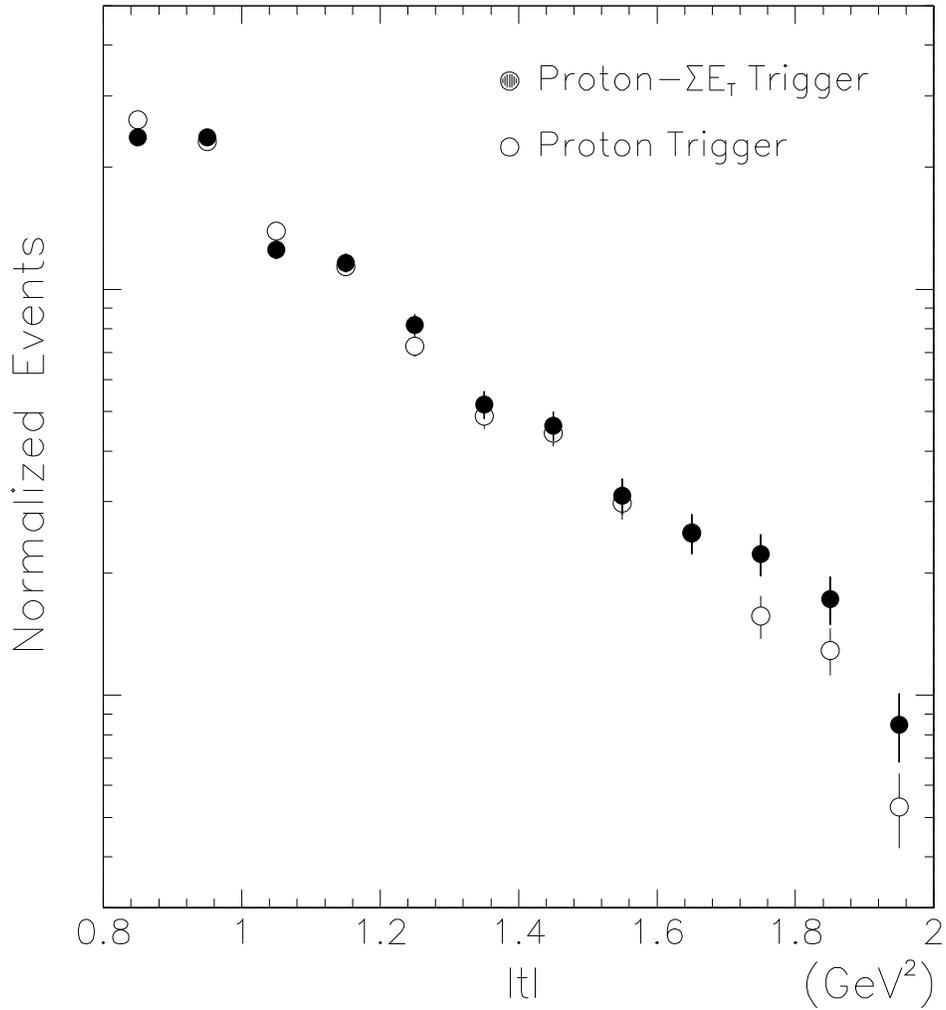,width=16cm}}
\end{center}
\caption[]{
Momentum transfer (\T ) dependence of the raw data samples for both ``DIF'' 
(inclusive-proton) and ``JET'' (proton-\sumet ) triggers, 
when $0.90 < \xp < 0.97$ and after (offline) rejection of pileup and halo
background\protect\cite{thesis}. The two distributions are normalized to
one another with an arbitrary scale.
}
\label{fig:traw}
\end{figure}

Figure~\ref{fig:traw} compares the uncorrected momentum-transfer distributions 
of data samples from the two triggers, with $0.90 < \xp < 0.97$ and 
with offline pileup and halo cleanup cuts imposed\cite{thesis}. 
We conclude that, to good approximation,
large-\sumet\ and low-\sumet\ events of React.~\ref{eq:ppdif} have the same 
\T -dependence (in our region of \T ). 

\begin{figure}
\begin{center}
\mbox{\epsfig{file=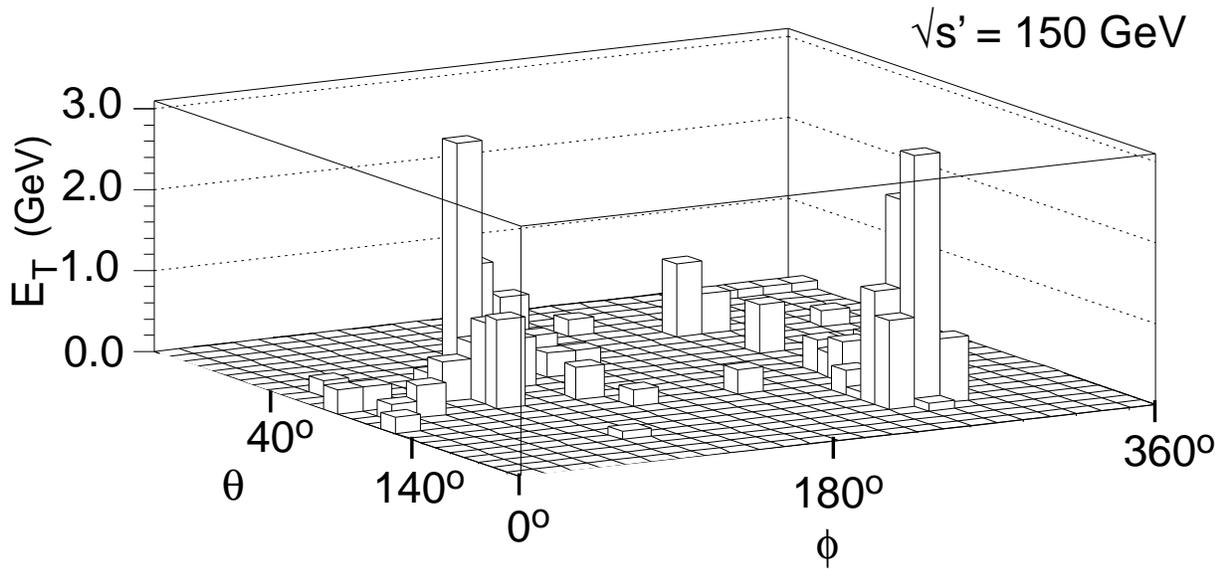,width=16cm}}
\end{center}
\caption[]{
A typical raw UA8 2-jet event display in the UA2 calorimeter:
cell energies in a $\theta$ vs. $\phi$ projection
(the complete event is shown).  
Each jet has $\etj > 8$~GeV.
The proton in this event had a measured  \xp\ = 0.94.
}
\label{fig:lego}
\end{figure}

Jet-finding was performed using UA2 calorimeter cell information, by requiring
that at least 8~GeV of transverse energy was deposited
within a cone of unit radius (in $\eta - \phi$)
around the direction of an initiator cell.
Figure~\ref{fig:lego} shows a display of a typical 2-jet event.
In this event, a recoil proton with  $\xp \simeq 0.94$ has carried away 
much of the initial state energy, leaving an effective interaction energy
$\rsp \simeq 150$~GeV.
The jets are clearly defined, with little underlying event background,
and are
separated by about $180^{\circ}$ in azimuthal angle, as expected for the hard
scattering of two partons 
(83\% of the 2-jet events have $\Delta \phi > 135^{\circ}$). 
The shapes and other characteristics of the 
jets were shown\cite{bonino,brandt} to agree with QCD Monte-Carlo predictions.
Table~\ref{tab:ratios} lists the numbers of 2-jet events in four bins
of \xp , where the jets satisfy a fiducial cut, $|\eta| < 2$, 
a coplanarity cut, $\Delta \phi > 135^{\circ}$ and are in a restricted
\T -range, 1.15-2.0~GeV$^2$.

We find that the fraction of triggered events with $\sumet > 22$ GeV
that contain jets is the same at low-\T\ and high-\T\ in our data.
The fraction is ($0.384 \pm 0.010$) for $0.9 < \T < 1.4$ GeV$^2$
and ($0.376 \pm 0.010$) for $1.7 < \T < 2.3$ GeV$^2$. 
Taking into account the observation that a \sumet\ selection itself does not
alter the \T -dependence, we conclude 
that {\it the \T -dependence of React.~\ref{eq:ppdif}
is the same with jets as without jets
over our \T -range}. We take this as a working assumption
for the analysis presented in this Letter and note that is consistent with
the hypothesis of factorization.
  
We define the parameter, ${\cal R}$, 
in a given $\Delta \xi$ bin in React.~\ref{eq:ppdif},
as the fraction of the total single diffractive cross section that exhibits
hard scattering.
Not only is the ${\cal R}$ ratio independent of \T\ within our \T -range,
but the acceptance corrections for protons or antiprotons 
as well as certain systematic uncertainties cancel.
\begin{equation}
\label{eq:rcalc}
{\cal R} \, \, = \, \, 
{{\Delta\sigma_{sd}^{jets}}\over{\Delta\sigma_{sd}^{total}}}
\, \, = \, \, 
{{N_j/({\cal L}_j\epsilon_j A_j)}\over{N_{sd}/({\cal L}_{sd}\epsilon_{sd})}}
\, \, = \, \, {{N_j}\over{N_{sd}}} \x {{1}\over{A_j}} 
\x {{{\cal L}_{sd}}\over{{\cal L}_j}} \x {{\epsilon_{sd}}\over{\epsilon_j}} 
\end{equation}
$N_j$ and $N_{sd}$ are the numbers of diffractive jet events and 
inclusive single diffractive
events, respectively (the 1989 data sample used in the present analysis
had a luminosity for the sample of jet events of ${\cal L}_j = 423$~nb$^{-1}$).
The efficiencies, 
$\epsilon _j = 0.50$ and $\epsilon _{sd} = 0.83$,
correct for good events which are
lost in the offline rejection of pileup and halo 
events\cite{ua8dif,ua8hard,thesis}.
$A_j$ is the jet acceptance\cite{thesis} 
for the events in the numerator and, 
for a hard gluonic \pom , is 0.44 at \xp\ = 0.91,
decreasing to 0.19 at \xp\ = 0.965. $A_j$ is 20\% larger for a 
hard $q\bar{q}$ \pom .

$A_j$ was calculated with a modified version 
of the PYTHIA 4.8 event generator\cite{pythia,pompyt}, in which
the \pom\ is defined as a beam particle, with gluonic or $q\bar{q}$
structure,
and a proton is the target
particle\footnote{We have since used POMPYT 2.6, which is based on
PYTHIA 5.7, to verify that our \pom\ structure conclusions are unaffected
by changing to the most recent proton structure functions.}. 
Hard \pom -proton interactions at a specific \rsp\ are calculated
for any assumed \pom\ structure function, 
using standard QCD parton-parton 
scattering matrix elements with initial and final state
radiation.
In PYTHIA, the minimum transverse momentum of the parton-parton
hard scattering, QTMIN, was set to 1/2 the desired jet threshold of 8 GeV,
in order to maximize the fraction of generated events that are useful for
the analysis without biasing the jet distribution.

JETSET 6.3\cite{jetset} 
was used to model the hadronization according to the
Lund string model\cite{lund}. 
The generated Monte Carlo events were then boosted
from the \pom -proton system to the laboratory frame where they were 
passed through the UA2 calorimeter simulation\cite{ua2mc}.
Finally, the simulated event sample was passed through
pattern recognition, jet-finding and selection software, identical
to that used in the processing of real data.

This procedure allows us to relate the number of events with two 
$\etj > 8$ GeV jets, to the 
events generated with scattered partons with $\pt > 8$ GeV.
Defining $A_j$ as the ratio of these numbers,  we follow a convention 
where the {\it scattered parton cross section} is quoted 
as the ``jet cross section'',
thus facilitating comparison with theoretical predictions.

Equation~\ref{eq:rcalc} is evaluated for diffractive
mass bins from 118 to 189 GeV  and the resulting
values of ${\cal R}$ are given in Table~\ref{tab:ratios}. 
${\cal R}$ is evaluated for both a hard gluonic and
a hard $q\bar{q}$ \pom , differing only in the $A_j$ value used, 
and is found to be in the range 0.0017 to 0.0028.
The absolute jet cross sections are given below. 

The dominant source of the systematic uncertainty in ${\cal R}$ (26\%) 
is the jet acceptance calculation,
to which three sources contribute equally and are combined in quadrature: 
uncertainties in ``tuning''
PYTHIA to describe the underlying events, the choice of the proton
structure function and the choice of
the minimum transverse momentum of the parton-parton scattering.
Imperfect agreement of the jet-finding yield between Monte-Carlo and data,
when the cone size and initiator energy of the jet-finding algorithm are 
changed, leads to an ``Algorithm" error (10\%).
The estimated uncertainty (10\%) on the ratio of the efficiency parameters 
is dominated by the correction for
pileup-rejected events (superimposed diffractive event with a minimum-bias
event) that contain a diffractive
event which alone has \sumet\ above the trigger threshold.
These components are added in quadrature to give a total 30\%.

We note one point concerning the ``super-hard" component in the 
data\cite{brandt}. 
These events, whose 2-jet longitudinal momentum component in the 
\pom -proton center-of-mass is larger than 0.7,
constitute about 30\% of the entire 2-jet sample.
Although the super-hard events are included in the $N_j$ of 
Table~\ref{tab:ratios}, the component is not explicitly
included in the calculation of $A_j$. 
Since the jet-acceptance is about 20\% larger for these events than
for the hard structure function used in the calculation of $A_j$,
the total effect on the values of ${\cal R}$ of Table~\ref{tab:ratios} is 
$\sim 6$\%. 
However, we ignore this, because our systematic uncertainty is 30\%.


\section{Phenomenology}
\label{method}

\indent

We assume factorization, as in Ref.~\cite{is}, such that the observed 
hard-scattering cross section in React.~\ref{eq:ppdif} is a product of 
the \pom -Flux-Factor\cite{dl_hard}, \flux , and
the cross section for \pom -proton hard scattering.

The QCD hard scattering takes place between a parton in the \pom\
and a parton in the proton or antiproton and
is calculated by POMPYT 2.6 with default settings:
\begin{equation}
\label{eq:sigpP}
f \x \sigma_{{\cal P}p}^{jets} \, \, = \, \, 
\int \di x_1 \di x_2 \di \hat{\T} \, \sum_{i,k} g(x_1) \,G_i(x_2,Q^2) 
\, \frac{\di \hat{\sigma}_i^k}{\di \hat{\T}}.
\end{equation}
$\sigma_{{\cal P}p}^{jets}$ is the hard scattering cross section if
the momentum sum rule is valid for the \pom .
$x_1$ is the momentum fraction of a parton in the \pom\
with effective structure $x g(x) = f \x 6 x (1-x)^1 $, where
$f \neq 1.0$ denotes a violation of the momentum sum rule.
$x_2$ is the momentum fraction of a parton
in a proton with CTEQ2L structure function, $G_i(x_2,Q^2)$. 
The cross section is based on the standard QCD matrix elements,
${\di \hat{\sigma}_i^k}/{\di \hat{\T}}$, and 
the summations go over all possible parton--parton scattering subprocceses.
The scale, $Q^2$, of the proton structure is equated to $(\etj )^2$ and
$Q^2$ evolution of the \pom\ structure function is ignored, because it is
believed to be small in our \etj -range, as is any possible 
dependence on \T .
The leading order 
values\footnote{If we include NLO contributions, using an effective k factor,
the cross sections increase by 20-30\%.} of \sigpomjets\ for hard gluon and 
hard quark structures, are given in Table~\ref{tab:ratios}.

In the $\xi$-range in which non-\pom -exchange background is small enough
to be ignored (see below for a discussion of this point),
the hard-scattering and the total single diffractive single-arm 
cross sections in React.~\ref{eq:ppdif} can be written as:
\begin{equation}
\label{eq:sigdifjets}
{{d^2 \sigdifjets}\over{d \xi dt}} \, = \,  
\flux \, \x \, [f \x \sigpomjets (s')]
\end{equation}
\begin{equation}
\label{eq:sigdiftot}
{{d^2 \sigdiftot }\over{d \xi dt}} \, = \,  \flux \, \x \, \sigpomtot (s'),
\end{equation}
The ratio of Eqs.~\ref{eq:sigdifjets} and \ref{eq:sigdiftot} gives us, 
on the left-hand side, the measured \T -independent 
${\cal R}$ parameter defined in Sec.~\ref{cross}. 
The Flux-Factor cancels out on the right-hand side 
and we have Eq.~2 of Ref.~\cite{is}:
\begin{equation}
\label{eq:R}
{\cal R} (s') \, \, = \, \, 
{{\Delta\sigma_{sd}^{jets}}\over{\Delta\sigma_{sd}^{total}}}
\, \, = \, \, f \x{{\sigpomjets (s')}\over{\sigpomtot (s')}}. 
\end{equation}

Previously\cite{bonino}, we used the simple assumption, $f=1$. 
We also assumed a constant $\sigpomtot = 2.3$~mb, based on triple-Regge 
analyses\cite{fieldfox,kaidalov,cool,berger} of single diffractive data.
Now, however, we wish to determine $f$ from experiment
and allow \sigpomtot\ to have a proper Regge dependence on $s'$.
Our current analysis is carried out with the following steps:
\begin{itemize}
\item From Eq.~\ref{eq:R}, it is seen that the measurements of ${\cal R}$ 
({\it appropriately background corrected}) 
and calculations of $\sigpomjets (s')$ 
in Table~\ref{tab:ratios} permit the determination of the ratio, 
$f / \sigpomtot (s')$. 
\item Fitting Eq.~\ref{eq:sigdiftot} to inclusive single diffractive
data permits the \pom -proton total cross section, $\sigpomtot (s')$,
as well as parameters of the \pom\ flux factor, \flux , to be determined. 
This step of the analysis is made using much higher statistics, and 
data at different energies, which is necessary to determine \flux . 
Theoretical uncertainty in the value of the overall normalization constant,
$K$, in \flux\ means that only the product 
$K \sigpomtot (s')$ can be uniquely determined. 
\item The product of $f / \sigpomtot (s')$ and $K \sigpomtot (s')$
yields the quantity, $fK$, which can be directly used to make predictions with 
Eq.~\ref{eq:sigdifjets} (providing \flux\ is known).
Furthermore, the simplest factorization assumptions imply that $fK$
should be independent of both $s'$ and $s$
(see Sec.~\ref{results} for a further discussion of this point).
\end{itemize}

In a separate article\cite{ua8dif}, our collaboration has reported a complete 
analysis of inclusive single diffraction. 
Combined fits of Eq.~\ref{eq:sigdiftot} were made to UA8 and 
lower energy ISR data\cite{albrow} ($s$ = 551 and 930~GeV$^2$)
in the momentum transfer range, $0.15 < \T < 2.0$~GeV$^2$
with the following forms of \flux\ and $\sigpomtot (s')$:
\begin{equation}
\label{eq:DIF}
{{d^2 \sigdiftot}\over{d \xi dt}} 
\, = \,  \flux \x \sigpomtot (s') 
\, = \, [K \, F_1(t)^2  \, e^{bt} \, \xi^{1-2\alpha(t)}] 
\, \x \, \sigma_0 [(s')^{0.10} \, + \, \rm R \, \it (s')^{-0.32}]
\end{equation}
%
$F_1(t)^2$ is the standard Donnachie-Landshoff form-factor\cite{dl_dif}.
It is found that the \pom\ Regge trajectory  requires a quadratic 
term such that, $\alpha (t) = 1.10 + 0.25 t + \alpha '' t^2$, 
with $\alpha '' = 0.079 \pm 0.012$~GeV$^{-4}$.
The factor, $e^{bt}$, compensates for the effect that 
the quadratic term has on the normalization\footnote{Donnachie 
\& Landshoff\cite{dl_dif} suggest that 
\sigpomtot\ may also depend on momentum transfer, \T . We ignore that
possibility in this paper, but note that any such dependence may be 
absorbed in this e$^{bt}$ factor.}. $s'$ has units of GeV$^2$.

It has been found in Ref.~\cite{ua8dif} that a \sigpomtot\ with only one term 
is inadequate to understand the existing single diffractive data.
Thus, in analogy with all total hadronic cross 
sections\cite{dl_tot}, \sigpomtot\ is written with two 
components. 
The first term is due to \pom -exchange and dominates at large $s'$.
The second term is due to C=+1 ($a/f_2$) Reggeon exchange 
and dominates at small $s'$. 
The exponents in the two terms of \sigpomtot\ are from 
Refs.~\cite{dino2,cudell}.
R is a free parameter in the fits.
It may be noted that Eq.~\ref{eq:DIF} is equivalent to the 
\pom -\pom -\pom\ and \pom -\pom -Reggeon terms in a triple-Regge
expansion (see, for example, Ref.~\cite{fieldfox}).

\begin{figure}
\begin{center}
\mbox{\epsfig{file=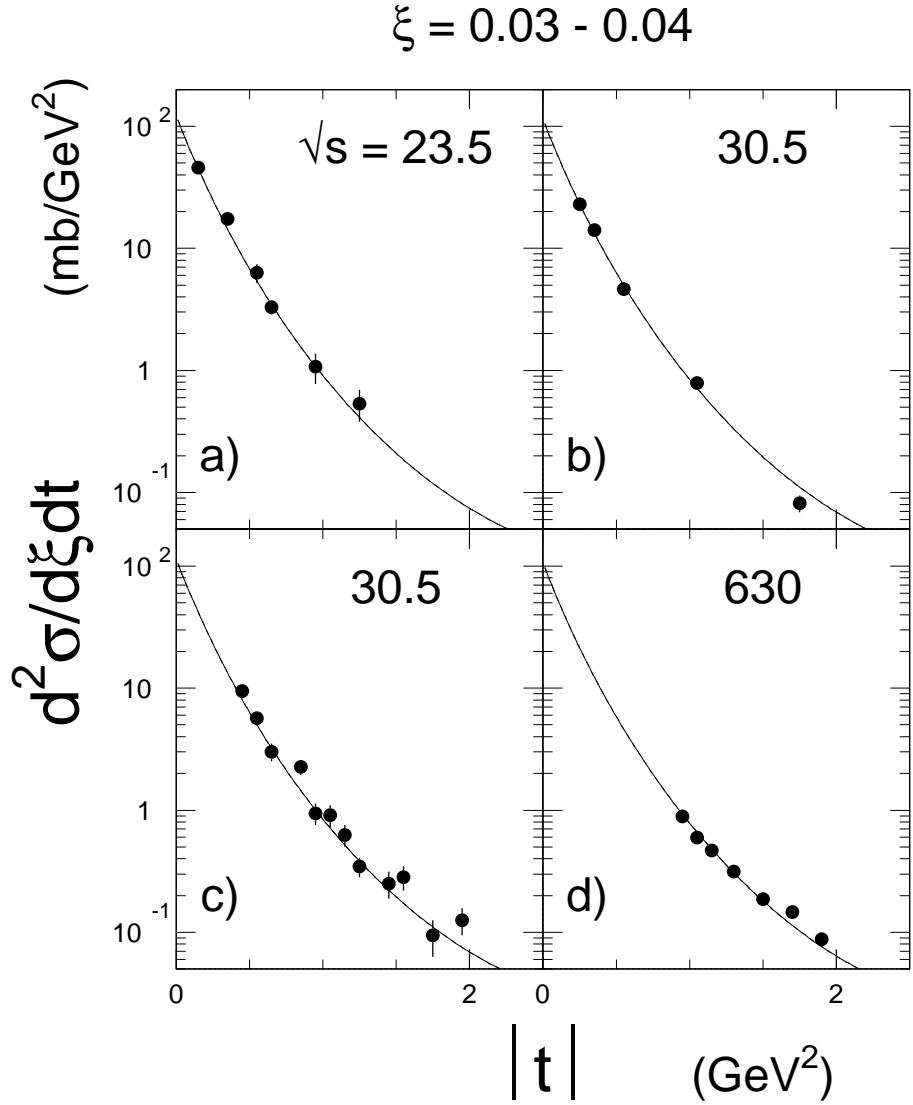,width=13cm}}
\end{center}
\caption[]{
Differential cross section, \dsig , vs \T , for three ISR 
measurements\protect\cite{albrow}
and UA8\protect\cite{ua8dif}. 
Points are averages of data in the $\xi$-range 0.03--0.04.
The curves correspond to Fit ``A"' in Table~\ref{tab:chisqbknd} evaluated
at $\xi = 0.035$. 
It is shown in Ref.~\protect\cite{ua8dif} that the relative normalizations of 
these data sets directly reflect the 
$s'$-dependence of \sigpomtot .
}
\label{fig:dsdt035}
\end{figure}

Table~\ref{tab:chisqbknd} shows the results\cite{ua8dif} 
for two of the various fits of Eq.~\ref{eq:DIF} to the data. 
Fit ``A" was made in the low-background region,
$0.03 < \xi < 0.04$, and the small residual background ($\sim 15\%$) was 
ignored. Fit ``A" is plotted in Fig.~\ref{fig:dsdt035} 
superimposed on the data in that $\xi$-bin and is seen to describe the
data quite well. 
Fit ``D" is made to data in the larger range $0.03 < \xi < 0.09$,
with a  conventional background term\cite{otherback}, 
$A \xi^1 e^{ct} $, added to Eq.~\ref{eq:DIF},
where $A$ and $c$ are different for the ISR data and the UA8 data.
The two types of fits are self-consistent.

There are several noteworthy results. First, we find:
\[ K\sigma_0 \, = \, 0.72 \pm 0.10 \rm \, \, \, mb \, GeV ^{-2} \]
which, if factorization is valid, provides 
a normalization to all diffractive processes.
Second, the value found for R (4.0 $\pm$ 0.6) is close
to that found in the fits to 
$pp$ and $p\ap$ total cross sections\cite{cudell,dino2}, illustrating that
the relative strengths of \pom -exchange and $a/f_2$ exchange in
\pom -proton scattering are similar to that found in $pp$ scattering.

We note that there is an implicit systematic uncertainty in the above
value of $K\sigma_0$ due to the choice of exponents of \sigpomtot\
in Eq.~\ref{eq:DIF} (see Refs.~\cite{dl_tot,dino2,cudell}). 
However, as discussed below, this particular
uncertainty cancels out when the product of $K\sigma_0$ and $f/\sigma_0$
is taken.


\section{Results}
\label{results}

\indent

\begin{figure}
\begin{center}
\mbox{\epsfig{file=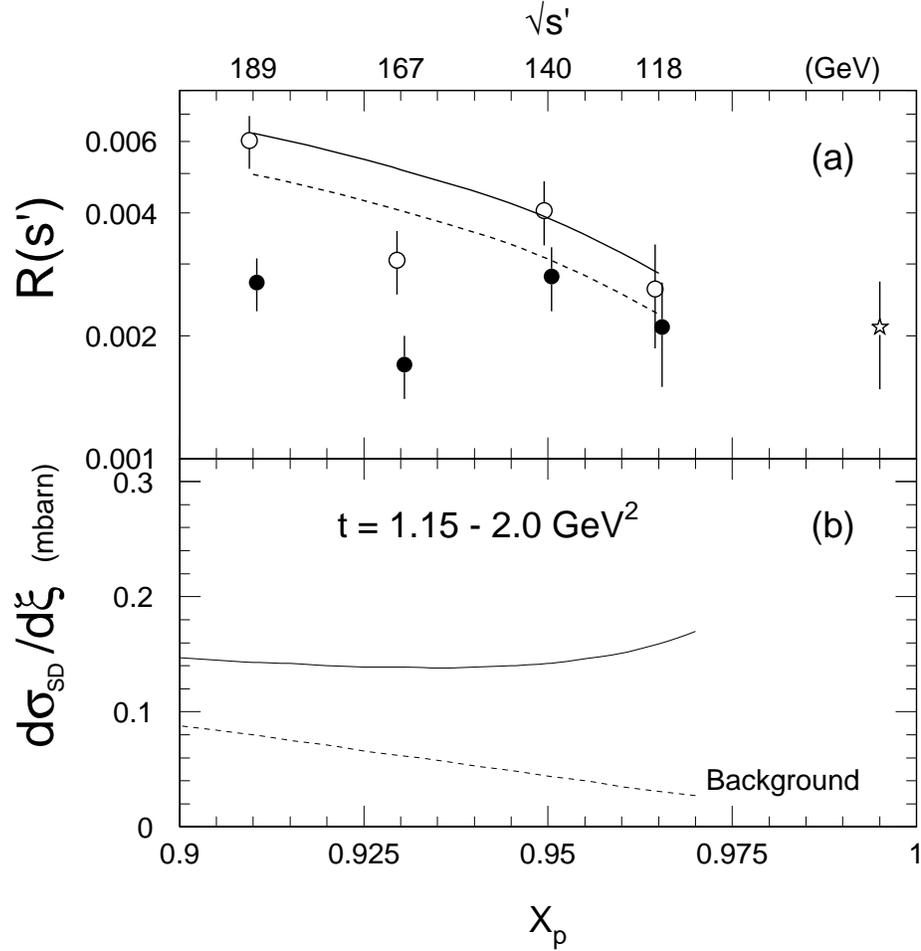,width=14cm}}
\end{center}
\caption[]{
(a) Measured cross-section ratio, ${\cal R}$ (solid points)
for a hard gluonic \pom , 
vs. proton momentum fraction and diffractive mass.
The star point on the right-hand-side shows the systematic 
uncertainty.
As explained in the text, the open points contain a correction for
non-\pom -exchange background in the denominator of ${\cal R}$.
The solid curve, normalized to the two right-hand points, is a prediction
discussed in the text. The dashed curve is the same, but normalized
to all four open points; 
(b) Solid curve is a fit to the measured differential
cross section\protect\cite{ua8dif}, 
$d\sigma _{sd} / d \xi$, for inclusive single diffraction.
The dashed curve is the fitted non-\pom -exchange background in
the observed cross section.
}
\label{fig:csec}
\end{figure}

We now use the techniques of the previous section to determine $fK$ and
the absolute cross section for jet production in React.~\ref{eq:ppdif}.
The solid points in Fig.~\ref{fig:csec}(a) are the experimental
${\cal R}$ ratios from Table~\ref{tab:ratios}, before corrections for 
non-\pom -exchange background.
To correct ${\cal R}$ for the background in its denominator, we divide it by 
the fraction of the single diffractive signal which is 
\pom -exchange\cite{ua8dif}, $\rm{{s}\over{s+b}}$, given
in Table~\ref{tab:sigmas}.
The results are plotted as the open points in Fig.~\ref{fig:csec}. 
We discuss below the possible contribution of background to the numerator.

Table~\ref{tab:sigmas} contains the fitted single diffraction cross 
section\cite{ua8dif}, \dsig , which has been
integrated over \T\ from 1.15-2.0~GeV$^2$, $\rm{{d\sigma_{sd}}\over{d\xi}}$. 
This quantity is multiplied by the ${\cal R}$ ratios 
in Table~\ref{tab:ratios} to find the absolute jet cross sections
given in Table~\ref{tab:sigmas}.
$\rm{{d\sigma_{sd}}\over{d\xi}}$ and the background contribution to
single diffraction, b, are plotted in Fig.~\ref{fig:csec}(b).

A prediction for the $s'$-dependence of the ${\cal R}$ ratio
in Fig.~\ref{fig:csec}(a) can be made using Eq.~\ref{eq:R},
with \sigpomtot\ replaced by the two-component version shown in 
Eq.~\ref{eq:DIF}.
The solid curve in Fig.~\ref{fig:csec}(a) shows this quantity:
\begin{equation}
\label{eq:Rp}
{\cal R} (s') \, \, = \, \, 
{ {\sigpomjets (s')}\over{(s')^{0.10} + 4 \, (s')^{-0.32}} } 
\, \x \, {{f}\over{\sigma_0 }}
\end{equation}
normalized to the two open points at largest \xp , where the background 
corrections are smallest. This yields a  fitted $f / \sigma_0$ value
of 0.532 $\pm$ 0.081(stat) $\pm$ 0.160(sys)~~~mb$^{-1}$.
The uncertainty from the choice of the exponents used is the same as
mentioned above for the determination of $K\sigma_0$, but now appears in the
denominator, so that there is a cancellation when the product is taken
to arrive at the final $fK$ values.

At this point, we note that background in the numerator of ${\cal R}$, 
jet events from ($\rm q \bar{q}$) Reggeon exchange, has been neglected.
The fact that the two open points at smaller \xp\ in Fig.~\ref{fig:csec} do 
not lie above the fitted solid curve signifies that such non-\pom -exchange
background is insignificant in the numerator. 
This may be understood by noting that the calculated ($\rm q \bar{q}$) jet
cross sections in Table~\ref{tab:ratios} are a factor 2.3 times smaller
than their gluonic counterparts. Furthermore, the Reggeon
flux factor is likely to be smaller than \flux .

Based on this argument, an improved determination of $f/\sigma_0$ 
should be possible by fitting Eq.~\ref{eq:Rp} to all four open
points in Fig.~\ref{fig:csec}. This fit is shown as the dashed curve in the
figure and yields:
\vspace{1mm}
\begin{tabbing}
\hspace{2.5cm}\=$f/\sigma_0$ = 
0.422 $\pm$ 0.039(stat) $\pm$ 0.127(sys)~~~mb$^{-1}$ \, \,   gluonic-\pom ,  \\
\hspace{2.5cm}\=$f/\sigma_0$ =
0.784 $\pm$ 0.072(stat) $\pm$ 0.235(sys)~~~mb$^{-1}$ \, \,   $q\bar{q}$-\pom .
\end{tabbing}
These values are only about one statistical standard deviation lower than
those obtained from fitting to the two points with largest \xp .

Multiplying these values of $f/\sigma _0$ by 
the above value, $K\sigma_0$ = 0.72, yields:
\vspace{1mm}
\begin{tabbing}
\hspace{2.5cm}\=$fK$ = 
0.304 $\pm$ 0.050(stat) $\pm$ 0.091(sys)~~~GeV$^{-2}$ \, \,   gluonic-\pom ,  \\
\>$fK$ = 
0.564 $\pm$ 0.094(stat) $\pm$ 0.169(sys)~~~GeV$^{-2}$ \, \,   $q\bar{q}$-\pom .
\end{tabbing}
With a dominant (80--90\%) gluonic component in the \pom\ reported by
the H1 Collaboration\cite{h1gluonic}, our gluonic value of
$fK$ ($0.30 \pm 0.10$) can be compared with corresponding
(jet cross section) measurements of $fK$ reported by 
ZEUS\cite{zeus} ($0.37 \pm 0.15$)
and CDF\cite{cdfdijet} ($0.11 \pm 0.02$).

If the \pom\ were like a real particle, the Donnachie-Landshoff value,
$K$ = 0.78~GeV$^{-2}$, 
is thought to be ``the only reasonable normalization of the
Flux-Factor"\cite{peter} and the momentum sum rule might be true ($f$ = 1.0).
We find however that, if $K$ has this value,
$f = 0.39$ for a gluonic-\pom , while for a $q\bar{q}$-\pom , $f$ = 0.72. 

With our determinations of $fK$ and \flux, 
hard diffraction cross section 
predictions may be calculated for Reacts.~\ref{eq:ppdif} and \ref{eq:epdif}.
It is interesting to note that the curve in Fig.~\ref{fig:csec}(a) 
is a prediction for the measured ratio, ${\cal R} (s')$, 
at any $s$-value in React.~\ref{eq:ppdif}, providing the $\rsp$-scale is used,
and background is taken into account.
Comparisons of these predictions with data samples 
from other experiments
will test the basic assumption of factorization used in our analysis.

Not discussed in this Letter is the issue of saturation of
\flux\ at high energies, which Goulianos\cite{dino} points out is required if
the triple-Regge prediction of
\sigdiftot\ is to agree with experiment and not violate unitarity.
We mention this here because Ref.~\cite{dino} proposes that saturation be
achieved by having $K$ decrease with increasing energy, $s$. 
However, Ref.~\cite{ua8dif} shows that the observed 
$s$-dependence of \dsig\ at fixed $\xi$ and \T\ is inconsistent with such
an $s$-dependent $K$, but is in good agreement with a constant
(i.e. $s$-independent) $K$ and a \sigpomtot\ with 2-components, as discussed
above. An alternate solution to the saturation of \flux\ at high energies has 
been proposed\cite{erhanschlein}, in which $K$ is $s$-independent and 
\flux\ is damped at small values of $\xi$ and $t$.

\section*{Acknowledgements}

We remain grateful to the UA2 collaboration, without whose 
cooperation these measurements would not have been possible,
and to the CERN administration for their support. 
We particularly wish to thank Sandy Donnachie for his strong early support
of this experiment.
We are indebted to John Collins, Gunnar Ingelman, 
and Peter Landshoff for helpful discussions.

\pagebreak

\clearpage

\begin{table}
\centering
\begin{tabular}{||c|c|c||c|c||c|c||}                         
\hline
 & & &\multicolumn{2}{|c||}{{$\boldmath \bf \rm  gluonic -   \bf \pom$ }} 
     &\multicolumn{2}{|c||}{{$\boldmath \bf \rm  q \bar{q} - \bf \pom$ }} \\
 & & &\multicolumn{2}{|c||}{ } &\multicolumn{2}{|c||}{ } \\
 $\xi$    &$\sqrt{s'}$    &$N_j $  
&${\cal R} =  {{\Delta\sigma_{sd}^{jets}}\over{\Delta\sigma_{sd}^{total}}}$
 &\sigpomjets\      
&${\cal R} =  {{\Delta\sigma_{sd}^{jets}}\over{\Delta\sigma_{sd}^{total}}}$
 &\sigpomjets\   \\     
 &     &      &         &     &      &     \\
 &GeV &    &$\times 10^{-3}$ &mb &$\times 10^{-3}$ &mb \\ 
\hline
.03-.04 &118 & 11 &$2.1 \pm 0.6$ &0.0149 &$1.8 \pm 0.5$ &0.0064 \\ 
.04-.06 &140 & 35 &$2.8 \pm 0.5$ &0.0209 &$2.3 \pm 0.4$ &0.0090 \\ 
.06-.08 &167 & 25 &$1.7 \pm 0.3$ &0.0282 &$1.4 \pm 0.3$ &0.0121 \\ 
.08-.10 &189 & 39 &$2.7 \pm 0.4$ &0.0353 &$2.3 \pm 0.3$ &0.0149 \\ \hline
\end{tabular}
\caption[]{
Numbers of 2-jet events, cross section ratios
corrected to scattered partons with $\pt > 8$~GeV, as explained in the text,
and calculated values of \sigpomjets\ for the same conditions. 
The ratios are for data in the momentum-transfer range, 1.15-2.0~GeV$^2$.
}
\label{tab:ratios}
\end{table}

\begin{table}
\centering
\begin{tabular}{|l r|c|c|} 
\hline 
                   &   &Fit ``A"     &Fit ``D"    \\
$\xi$-range        &   &0.03-0.04    &0.03-0.09   \\
\hline
$\chi ^2$          &   &65           &393         \\
No. of Data points &   &48           &292         \\
$\chi ^2$/degree of freedom &   &1.5          &1.4   	      \\
$K \sigma _0$& mb GeV$^{-2}$    &0.67 $\pm$0.08      &0.72 $\pm$0.10  \\
$\alpha ''$  &GeV$^{-4}$        &0.078$\pm$0.013     &0.079$\pm$0.012\\
$b$          &GeV$^{-2}$        &0.88 $\pm$0.19      &1.08 $\pm$0.20 \\
R            &                  &5.0  $\pm$0.6       &4.0  $\pm$0.6   \\
\hline 
$A$(UA8)          & mb GeV$^{-2}$ &--           &25$\pm$7     \\
$A$(551)          & mb GeV$^{-2}$ &--           &280$\pm$30   \\
$A$(930)          & mb GeV$^{-2}$ &--           &226$\pm$21   \\
$c$(UA8)          & GeV$^{-2}$     &--           &2.1$\pm$0.2   \\
$c$(ISR)          & GeV$^{-2}$     &--           &3.5$\pm$0.1   \\
\hline
\end{tabular}    
\caption[]{
Fit results\protect\cite{ua8dif} of Eq.~\ref{eq:DIF} to
experimental values of $d^2 \sigma / d\xi dt$~~(mb/GeV$^2$)
from UA8 and ISR\cite{albrow}.
Fit ``A" includes no background; 
Fit ``D" includes background of the form $A \xi ^1 e^{ct}$, $A$ and $c$ are
different for UA8 and ISR data. The bottom part of the table includes the
fitted parameters of the background.
}
\label{tab:chisqbknd}
\end{table}

\begin{table}
\centering
\begin{tabular}{|c|c|c|c|c|} 
\hline 
        &       &             &   &   \\
$\xi$   &\rsp\  &${{d\sigma_{sd}}\over{d\xi}}$      &$\rm {s}\over{s+b}$ 
                &${{d\sigma_{jets}}\over{d\xi}}$ \\
        &       &             &   &   \\
        &GeV    &mb           &   &$\mu$b \\
\hline
.03-.04 &118 &0.159 &0.81 &$0.33 \pm 0.10$ \\
.04-.06 &140 &0.142 &0.69 &$0.40 \pm 0.07$ \\
.06-.08 &167 &0.139 &0.55 &$0.24 \pm 0.04$ \\
.08-.10 &189 &0.143 &0.45 &$0.39 \pm 0.06$ \\
\hline
\end{tabular}    
\caption[]{
Single-diffractive (single arm) differential cross sections\cite{ua8dif} 
integrated over
the momentum transfer range, \T\ = 1.15-2.0~GeV$^2$.
Column 4 shows the fraction\cite{ua8dif} 
of the the single-diffractive cross section which is \pom -exchange.
The di-jet differential cross sections in Column 5,
for the same \T -range, are obtained by
multiplying Column 3 by the ${\cal R}$ values
for the gluonic-\pom\ in Table~\ref{tab:ratios}.
There is a 30\% systematic uncertainty on the jet cross sections.
}
\label{tab:sigmas}
\end{table}

\clearpage
 
\clearpage
 
\clearpage

\clearpage

\clearpage

\end{document}